\documentclass[preprint,5p]{elsarticle}
\usepackage{natbib}
\usepackage{graphics}

\usepackage{amssymb}
\usepackage{amsthm}
\usepackage{siunitx}

\usepackage{enumitem}

\usepackage{lineno}

\biboptions{sort&compress}

\journal{Nuclear Instruments and Methods in Physics Research A }

\begin{document}

\begin{frontmatter}



\title{Energy acceptance of the St.~George recoil separator}


\author[Add1,Add2]{Z.~Meisel\corref{cor1}}
\ead{meisel@ohio.edu}
\cortext[cor1]{Corresponding author}
\author[Add2]{M.T.~Moran}
\author[Add2,Add3]{G.~Gilardy}
\author[Add4,label2]{J.~Schmitt}
\fntext[label2]{Present Address: Department of Physics and Astronomy, Michigan State University, East Lansing, MI 48824, USA}
\author[Add2]{C.~Seymour}
\author[Add2]{M.~Couder}
\ead{mcouder@nd.edu}
\address[Add1]{Institute of Nuclear \& Particle Physics, Department of Physics \& Astronomy, Ohio University, Athens, OH 45701, USA}
\address[Add2]{Department of Physics,
Joint Institute for Nuclear Astrophysics, University of Notre Dame, Notre Dame, IN 46556, USA} 
\address[Add3]{Centre d'\'{E}tudes Nucl\'{e}aires de Bordeaux Gradignan, UMR 5797 CNRS/IN2P3 - Universit\'{e} de Bordeaux, 19 Chemin du Solarium, CS 10120, F-33175 Gradignan, France}
\address[Add4]{Department of Physics \& Astronomy, Clemson University, Clemson, SC 29634, USA}

\begin{abstract}
Radiative alpha-capture, $(\alpha,\gamma)$, reactions play a
critical role in nucleosynthesis and nuclear energy generation in a
variety of astrophysical environments. The St.~George recoil
separator at the University of
Notre Dame's Nuclear Science Laboratory was developed to measure
$(\alpha,\gamma)$ reactions in inverse kinematics via recoil detection 
in order to obtain nuclear reaction cross sections at the
low energies of astrophysical interest, while avoiding the
$\gamma$-background that plagues traditional measurement techniques.
Due to the $\gamma$ ray produced by the nuclear reaction at the target
location, recoil nuclei are produced with a variety of energies and
angles, all of
which must be accepted by St.~George in order to accurately determine
the reaction cross section. 
We demonstrate the energy acceptance of
the St.~George recoil separator using primary beams of helium,
hydrogen, neon, and oxygen, spanning the magnetic and electric rigidity
phase space populated by recoils of
anticipated $(\alpha,\gamma)$ reaction measurements. We find the performance of St.~George meets
the design specifications, demonstrating its suitability for
$(\alpha,\gamma)$ reaction measurements of astrophysical interest.

\end{abstract}


\end{frontmatter}


\section{Introduction}
\label{Introduction}

Precise $(\alpha,\gamma)$ reaction cross sections are crucial to
accurately model a variety of astrophysical phenomena~\cite{Brun15}, such as 
the production of the $s$-process
neutron-source nucleus
$^{22}$Ne~\cite{Kapp94}, the
carbon-to-oxygen abundance ratio of stellar cores~\cite{Wies12}, nucleosynthesis
in massive stars~\cite{Tur09}, the abundance of $^{44}$Ti produced in
core-collapse supernovae~\cite{Magk10}, and the nuclear reaction
sequence occurring in type-I x-ray
bursts~\cite{Cybu16}.
Owing to the relatively low
kinetic energies of nuclei in the relevant astrophysical conditions,
traditional nuclear reaction measurements in which the outgoing
$\gamma$ ray is measured often suffer from prohibitively high
$\gamma$-backgrounds and, for high $Q$-value reactions, complicated
$\gamma$-ray cascades~\cite{Caci09}. One approach which has been adopted to overcome this
difficulty employs inverse kinematics and a recoil separator, in which the
nuclear recoil produced in the $(\alpha,\gamma)$ reaction is
electromagnetically separated from unreacted beam
nuclei on the basis of their mass differences
and identified with various combinations of time-of-flight and
energy-loss measurements~\cite{Ruiz14}. Recoil--$\gamma$ coincidences
substantially improve the signal-to-noise compared to traditional
techniques and provide significantly reduced cross section
uncertainties~\cite{Aker13}.

In order to undertake $(\alpha,\gamma)$ reaction cross section
studies at astrophysically relevant energies for nuclei with nuclear
mass $A\leq40$, the St.~George recoil
separator at the University of Notre Dame's Nuclear Science
Laboratory (NSL) has been developed~\cite{Coud08}.
The recoil separator technique generally relies on using
inverse kinematics, where a heavy ion beam is impinged on a lighter
nuclear target; for $(\alpha,\gamma)$ reactions with St.~George, the 
HIPPO helium gas-jet produces the desired target~\cite{Kont12,Meis16}. Unreacted incident
beam particles and nuclear reaction recoils both exit the gas-jet
with similar momenta. For a typical
cross section of $1$~$\mu$b and a gas-jet density of
$10^{17}$~atoms/cm$^{2}$, the ratio for recoils to unreacted
beam nuclei is 1 part in $10^{13}$. In order to be effective,
a recoil separator must accept all (or most~\cite{Mate06}) of the
nuclear reaction recoils and have a mass separation sufficient to
produce a recoil/beam ratio allowing for practicable particle
identification.

Due to the prompt $\gamma$-ray emission, nuclear reaction
recoils leave the target with a range of energies and angles,
according to the reaction $Q$-value, $\gamma$-ray cascade, and
$\gamma$ emission angles~\cite{Ruiz14}. 
For St.~George, the `acceptance' is the angle-energy phase space of
nuclear reaction recoils that the separator transmits to the recoil
detection-plane with 100\% efficiency.
Accurate cross section measurements with
the recoil separator technique rely on a thorough characterization
of the recoil separator acceptance.

\begin{figure*}[ht]
\begin{center}
\includegraphics[width=1.5\columnwidth,angle=0]{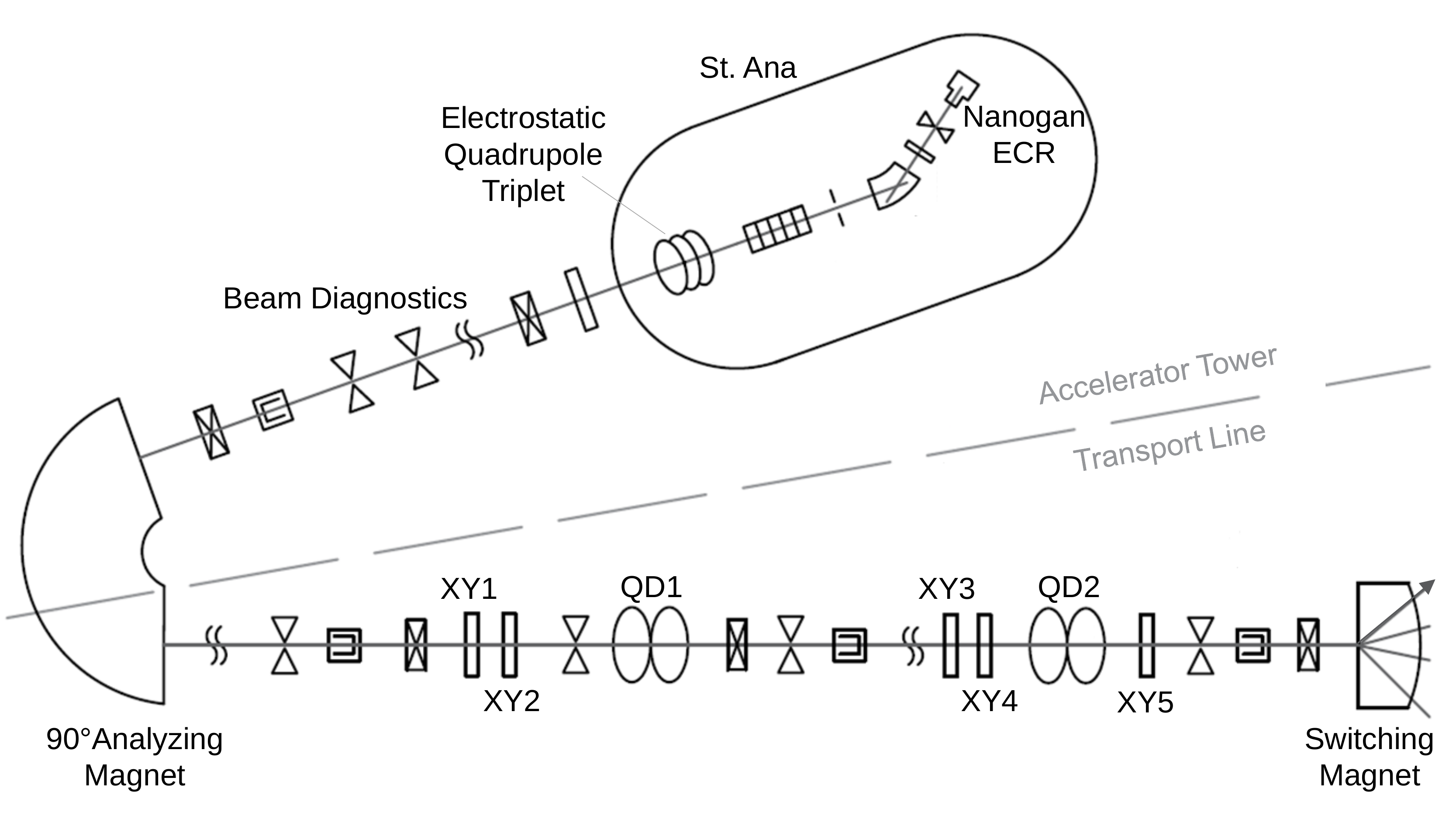}
\caption{
 Schematic of the St.~Ana accelerator and transport beam line to
 St.~George (The vertical accelerator beam line is folded here for
 convenience.).
 ``QD" correspond to quadrupole magnet doublets, which are
 used to focus the beam, while ``XY" correspond to steerer
 magnets, which are used to ensure the beam is transported down the
 ion optical centers of the quadrupoles via small corrections. 
\label{5UtransportSchematic}}
\end{center}
\end{figure*}

In a crucial first step of the St.~George recoil separator
commissioning process, we have determined the energy acceptance of
St.~George for a range of beam species, spanning the magnetic and
electric rigidity phase space anticipated for recoils of future
$(\alpha,\gamma)$ cross section measurements.
We discuss the
experimental set-up in Section~\ref{ExptSetup}, our energy
acceptance measurement method in Section~\ref{EnergyMethod}, and the
measurement results and discussion thereof in
Section~\ref{Results}
, prior to summarizing our work in
Section~\ref{Conclusions}.

\section{Experimental set-up}
\label{ExptSetup}

Stable ion beams are delivered to St.~George by the recently
installed 5~MV single-ended
Pelletron accelerator St.~Ana, from National Electrostatics
Corporation\footnote{http://www.pelletron.com}, at the NSL.
Ion beams produced by a Nanogan ECR ion source, from
Pantechnik\footnote{http://www.pantechnik.com},
are accelerated with St.~Ana, energy-selected by a 90$^{\circ}$
analyzing magnet, and transported down a $\approx$22~m
beam-line, which contains several quadrupole and 
steerer magnets, to St.~George
with beam intensities upwards of 100~$\mu$A. The 
configuration of the transport line at the time of this experiment, shown in Figure~\ref{5UtransportSchematic},
consisted of the horizontal X and vertical Y steerer magnet pair
XY$_{1}$ and XY$_{2}$
following the accelerator
momentum-analyzing magnet and preceding the quadrupole
doublet QD$_{1}$, followed by said quadrupole doublet, steerer
pair XY$_{3}$ and XY$_{4}$, the second quadrupole doublet QD$_{2}$,
the single steerer XY$_{5}$, a
dipole `switching' magnet, and (not shown) an XY steerer and a
quadrupole triplet. Beam position monitors, Faraday cups, and
fluorescing quartz
windows are employed at various locations along the transport line
to provide diagnostic information during beam tuning.

\begin{figure*}[ht]
\begin{center}
\includegraphics[width=1.5\columnwidth,angle=0]{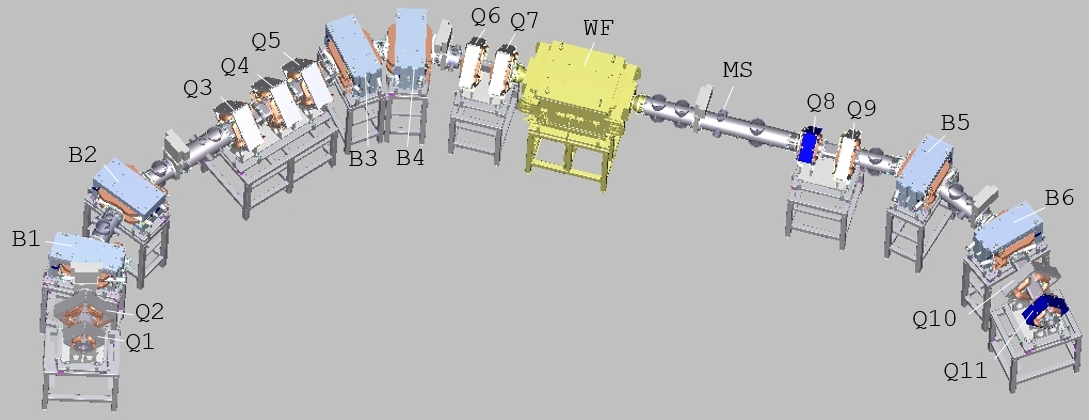}
\caption{
 Schematic of the St.~George recoil separator, where the labels
 correspond to ``B" for dipole magnets, ``Q" for quadrupole magnets,
 ``WF" for the Wien filter, and ``MS" for the mass slits.
 The target position is located
 upstream of Q1 and the recoil detection system (not shown) is
 downstream of Q11.
\label{StGeorgeSchematic}}
\end{center}
\end{figure*}

St.~George consists of eighteen ion optical elements: six dipole magnets, eleven quadrupole
magnets (four doublets and one triplet), and a Wien filter, as shown in Figure~\ref{StGeorgeSchematic}.
The last quadrupole, Q11, is
followed by an ion detection chamber in which ions can be identified
via total energy loss and time-of-flight. Several ports are available
along the St.~George beam line and on the magnets for
diagnostic equipment to assess the properties of transmitted ions.
For future $(\alpha,\gamma)$ reaction measurements, St.~George will
be preceded by the HIPPO gas-jet~\cite{Kont12}, which will serve
as a relatively uniform high-density reaction target~\cite{Meis16}.
For the acceptance measurements presented here, the primary beam from St.~Ana
was transmitted directly into St.~George, where two 2~mm diameter
collimators 20~cm apart were employed just prior to the
entrance of St.~George to ensure the beam entered St.~George on-axis.
The collimators had the added benefit of
limiting beam currents to a few $\mu A$ in order to
protect diagnostic equipment.

Diagnostic equipment along St.~George includes Faraday cups,
fluorescing quartz windows, and electrically insulated slits.
Faraday cups, used to monitor beam transmission, are located
following the first, second, and fifth dipole magnets, the Wien
filter, and the eleventh quadrupole magnet. Each Faraday cup is
equipped with a -300~V secondary electron suppression electrode to ensure
reliable current readings.
Electrically insulated slits with current readouts, used for the energy acceptance
measurements to monitor the beam width and centering, are located
before and after each dipole magnet, as well
as 1.59~m downstream of the Wien filter exit. These slits will be used
in future $(\alpha,\gamma)$ reaction measurements to improve
beam-rejection properties of St.~George. In particular, the slits
following the Wien filter, referred to as the `mass slits', are
located at the position optimized for
beam-rejection, which provides a horizontal
achromatic focus for recoil nuclei.
The slits and Faraday cups
each have a lower-limit for their current sensitivity of $\approx$1~pA.  The $0^{\circ}$ exit port of the
first, third, and fifth dipole magnets, as well as the exit port of
the recoil detection
chamber, are equipped with fluorescing quartz windows from
McMaster--Carr\footnote{http://www.mcmaster.com}, outside of
which (at atmosphere) is a camera. The quartz windows are used to
assess the beam shape at their respective locations and, more
importantly, are used to determine if the beam is being
transported along the ion-optical center of the quadrupole magnets
preceding that quartz. The quartz viewers also serve as a crude way
to determine beam transmission, as their fluorescence is visible at
relatively low beam currents.


\section{Energy acceptance measurement method}
\label{EnergyMethod}
In inverse kinematics, the products of an $(\alpha,\gamma)$ reaction
are emitted from the target position with the
same average momentum as the unreacted beam 
in a narrow cone, where the opening angle is primarily determined by
the reaction $Q$-value and center-of-mass energy (See Equation~13 of
Reference~\cite{Ruiz14}.). St.~George is
designed to accept and transmit all recoil nuclei of a single
charge-state 
emitted from the target with a maximum angle of $\pm$40~mrad
and energy spread from the central recoil energy of
$\pm$8\%~\cite{Coud08}. The maximum recoil energy deviations occur
when $\gamma$ rays are emitted at $0^{\circ}$ or $180^{\circ}$ in the
center of mass.

In this study, we mimicked the recoil energy spread
by employing a beam in the absence of a target in order to verify
that, at $0^{\circ}$ (i.e. for no angular deflection), 100\% of ions within $\pm$8\% of the energy
for which the separator was tuned were transmitted to the end of
St.~George. A similar technique was employed to determine the energy
acceptance of the ERNA recoil separator~\cite{Roga03,Roga99}.
Transmission through the recoil separator was determined by
comparing beam current measurements provided by the Faraday cups at
the target location and in the recoil detection plane, where equal
currents correspond to 100\% transmission. Various
charge states and beam energies were employed for ions of helium, hydrogen,
neon, and oxygen in order to determine the energy acceptance for
nuclei within the designed electric and magnetic
rigidity\footnote{For an ion of kinetic energy $T$ in MeV, mass $A$
in MeV, and charge $q$ in units of the electron charge,
$\rm{B}\rho=\frac{\sqrt{\text{\emph{T}}^{2}+2\text{\emph{TA}}}}{300\text{\emph{q}}}$~Tm and
$\rm{E}\rho=\frac{2\text{\emph{T}}}{\text{\emph{q}}}$~MV.}
limits of St.~George,
$\rm{E}\rho\leq5.7~\rm{MV}$ and
$0.1~\rm{Tm}\leq \rm{B}\rho\leq0.45~\rm{Tm}$, respectively~\cite{Coud08}.

The energy acceptance measurements for each species, defined by a
given $\rm{B}\rho$--$\rm{E}\rho$, consisted of two main procedures, which
each consisted of several steps. Summarized briefly, the first
procedure consisted of adjusting the magnetic elements prior to and within
St.~George to ensure that the ion beam was traveling down the ion
optical center of the separator when at the tune-energy. The second procedure
consisted of adjusting the focusing elements within St.~George to
maximize the energy range about the tune-energy within which all ions
would be transmitted through the separator. 
The second procedure was necessary to find the optimal tune about
the nominal ion optical settings~\cite{Coud08} provided by
calculations performed with the software {\tt COSY
Infinity}\footnote{http://www.bt.pa.msu.edu/index\_cosy.htm}~\cite{Maki06}.

\begin{figure}[ht]
\begin{center}
\includegraphics[width=0.4\columnwidth,angle=0]{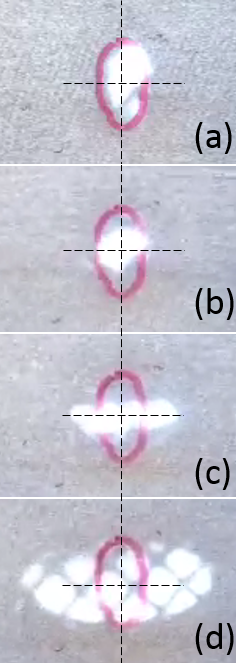}
\caption{
 Demonstration of a non-steering ion optical tune. (a)-(d)
 show images of the 
 B3 dipole magnet
 $0^{\circ}$ quartz viewer for a 1~MeV $\rm{H}^{+}$ beam, where the
 Q2 quadrupole magnet is scaled from its minimum field strength (a) up to
 $\approx0.13$~T,
 which is $\sim$50\% of its maximum value (d), where (b) and (c) are
 at arbitrary intermediate field strengths. The red oval and black crosses
 provide a reference for the size and center of the beam. The height
 and width of the
 red oval correspond to 
 $\approx$3~mm and $\approx$1.5~mm, respectively, on the quartz viewer.
\label{NonSteeringBeamSpot}}
\end{center}
\end{figure}

The first procedure, centering the ion beam within St.~George so
that quadrupole magnets only provided focusing and not beam steering, was
accomplished by performing the following steps:
\begin{enumerate}[label=1.\arabic*.]
 \item The accelerator magnetic and electrostatic elements and
 St.~Ana transport beam-line magnetic elements
 (See Figure~\ref{5UtransportSchematic}.) were adjusted to deliver
 an ion beam of the chosen species and energy with a current of
 several hundred nA to the St.~George target location. The
 accelerator and transport line elements were adjusted to
 send beam through the two collimators at the target location.
 \item With no field in the B1 dipole magnet\footnote{We note that
 residual magnetic fields on the dipoles do not impact our methods,
 as the effect would only be a horizontal translation of the beam
 spot on a quartz viewer.}, the beam was
 impinged on the B1 $0^{\circ}$ quartz viewer.
 The Q1 and Q2
 quadrupole magnets of St.~George separately had their field
 strengths adjusted from the minimum to $>50$\% of the maximum field
 strength. The image of the beam-spot on the B1 $0^{\circ}$
 exit-port fluorescing-quartz was monitored for shifts in the spot
 centroid. Any deviation from a `non-steering' beam, shown in
 Figure~\ref{NonSteeringBeamSpot}, was
 corrected by adjusting the last few steering magnets of the St.~Ana
 transport line.
 \item The B1 and B2 dipole magnets were adjusted to their
 nominal fields for the species of interest at the tune-energy,
 sending the beam to the B3 $0^{\circ}$ exit-port quartz-viewer,
 with no field on the B3 dipole. The
 correction process of the previous step was repeated. Though this step is seemingly
 redundant, the longer distance of travel of the beam from Q1 and Q2
 to this quartz provided a higher sensitivity to the presence of
 steering in the ion optics. This step could not be repeated for
 the $0^{\circ}$ exit port of the B5 dipole magnet since the ion
 beam was too defocused at that location without using quadrupole
 magnets
 Q3--Q9, a necessary condition to test for beam steering.
 \item With fields on Q1 and Q2 set to provide a narrow waist at the B3
 $0^{\circ}$ exit-port quartz-viewer, the B1 and B2 fields were
 adjusted to minimize steering in the quadrupole magnets Q3--Q5.
 These adjustments were generally within $\lesssim0.3$\% of the
 calculated optimal field setting. A tune with simultaneously
 non-steering Q1--Q2 and Q3--Q5 was not achievable, likely due to
 a minor misalignment of the quadrupole magnets. However, the shift
 of the beam spot position from steering
 was minimized to a vertical deflection of $\lesssim2.5$~mm on the quartz viewer, which
corresponds to a maximum deflection of $\lesssim2$~mrad by the Q5
 quadrupole magnet at maximum field strength. Given the Q5 field
 gradient, the observed deflection indicates the vertical
 misalignment of the Q3--Q5 quadrupoles is $\lesssim1$~mm with
 respect to the Q1--Q2.
 \item Quadrupole magnets Q1--Q5 and dipole magnets B3--B4 were adjusted to their nominal
 fields for the species of interest at the tune-energy, sending the
 beam to the B5 $0^{\circ}$ exit-port quartz-viewer, with no field
 on the B5 dipole. A small electric field was applied to the Wien
 filter to offset the residual magnetic field.
 The B3 and B4 fields were adjusted to achieve the
 non-steering condition on the B5 $0^{\circ}$ exit-port
 quartz-viewer for the Q6--Q7 quadrupole magnets.
\begin{figure*}[ht]
\begin{center}
\includegraphics[width=2.\columnwidth,angle=0]{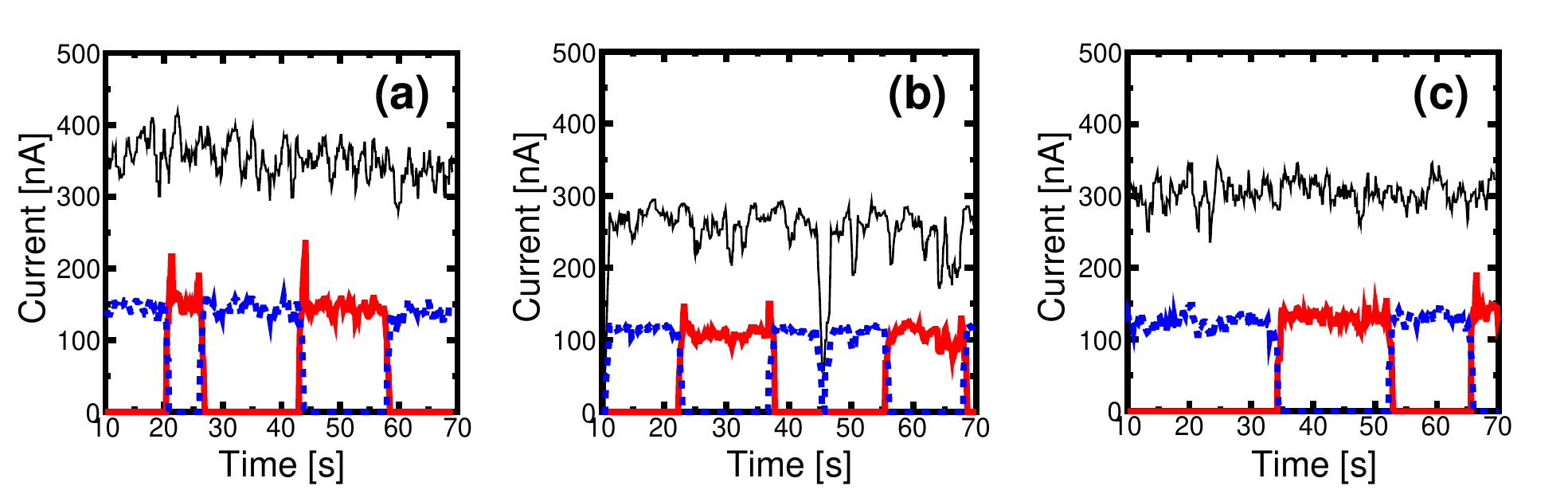}
\caption{
  (color online.) Demonstration of 100\% beam transmission through St.~George for $^{16}\rm{O}^{2+}$
  ions with the recoil separator tuned for $\rm{B}\rho=0.421$~Tm,
  $\rm{E}\rho=2.0$~MV (2~MeV ions, in this case.). Figures (a), (b),
  and (c) show the beam current read at the target location (thick
  red lines), detector plane (thick blue-dashed lines), and collimator
  upstream of St.~George used to monitor the incoming beam-current
  stability (thin black lines), for
  2~MeV (a), 2.16~MeV (b), and 1.84~MeV  (c) ions.
\label{Transmission16O}}
\end{center}
\end{figure*}
 \item The Wien filter magnetic and electric fields, as well as the
 magnetic field clamps, were set to
 their nominal values for the species of interest at the
 tune-energy. 
 The Wien filter magnetic field was then adjusted to
 achieve the non-steering condition for the Q8--Q9 quadrupole
 magnets on the B5 $0^{\circ}$ exit-port quartz-viewer.
 \item The electrically insulated slits with charge readouts located
 1.59~m downstream of the Wien filter, the so-called `mass
 slits', were then closed to the point of reading $\lesssim1$~pA on
 the left and right mass slits. This provides a measurement of the
 beam width, which we can use to estimate the anticipated
 beam-rejection of the separator.
 \item The B5 and B6 dipole magnets were then set to and adjusted
 about their nominal field settings for the species of interest at
 the tune-energy until the non-steering condition was achieved for
 the quadrupole magnets Q10--Q11 on the quartz-viewer located on the
 $0^{\circ}$ exit port of the recoil detection chamber.
\end{enumerate}

The second procedure, maximizing the energy acceptance of St.~George,
was accomplished by performing the following steps:
\begin{enumerate}[label=2.\arabic*.]
 \item Transmission was checked by monitoring the beam current on a
 collimator upstream of the target location and monitoring the beam
 current on Faraday cups at the target and recoil detection locations,
 while extending and retracting the target location Faraday cup (See
			Figure~\ref{Transmission16O}.). For the calculated ion optical tune for the species of interest,
 we did not find 100\% beam transmission from the target location to
 the recoil detector plane. Therefore, the quadrupole magnets Q1--Q11
 required adjustment to achieve 100\% transmission. With all other
 quadrupole magnets set to their nominal field setting, each
 quadrupole was individually adjusted in order to maximize the beam
 transmission to the recoil detector plane. Then, for each of the $N$ quadrupole magnets
 which improved beam transmission upon adjustment, the field was
 adjusted from the nominally calculated value by $1/N$ times the
 individual adjustment required for that quadrupole magnet to
 improve beam transmission. This na\"{i}ve scaling procedure
 resulted in 100\% beam transmission (See
 Figure~\ref{Transmission16O}a.).
 \item Having achieved 100\% beam transmission for the tune-energy,
 the St.~Ana terminal voltage was adjusted so that the beam would have an
 energy -8\% below the tune-energy,
 e.g. to 1.84~MeV from a tune-energy of 2~MeV. In general, 100\%
 transmission was not achieved. If beam current was lost on the mass
 slits, quadrupole magnets Q1--Q6
 were adjusted as
 described in the previous step to optimize beam transmission.
 Once beam current was removed from the mass slits, quadrupole
 magnets Q8--Q11 were adjusted as described in the previous step to
 optimize beam transmission.
 \item Beam transmission was checked again with the accelerator
 adjusted back to the tune-energy. If 100\% beam transmission was
 not achieved, the procedure from the previous step was repeated.
 This and the previous step were performed iteratively until 100\%
 beam transmission was obtained for both the tune-energy and the
 -8\% $\frac{\Delta E}{E}$ energy.
 \item Having achieved 100\% transmission for the tune-energy and
 -8\% below the tune-energy, the St.~Ana terminal voltage was adjusted so
 that the beam would have an energy +8\% above the tune-energy, e.g.
 to 2.16~MeV from a tune-energy of 2~MeV. If 100\% transmission was
 not achieved, adjustments needed to be made to Q1--Q11. These
 adjustments, along with a re-verification of transmission at the
 tune-energy and -8\% below the tune-energy, were performed in an
 iterative fashion, as described in the previous step. In general,
 we found the ion optical tune needed little to no adjustment to
 achieve 100\% transmission at +8\% above the tune-energy once 100\%
 transmission was achieved for the tune-energy and -8\% below the
 tune-energy.
\end{enumerate}

We note that the second of the two aforementioned procedures was
greatly expedited for ensuing energy acceptance measurements, as the
the suspected optimum values for the nominal tune were improved with time.
For instance, for the third successful energy acceptance measurement
only a single adjustment of quadrupole magnets Q10--Q11 was required
in order to achieve $\pm8$\% energy acceptance.
 
Figure~\ref{Transmission16O} demonstrates $\pm8$\% energy-acceptance
for a 2~MeV $^{16}\rm{O}^{2+}$ beam. The fluctuations observed in
beam intensity are thought to be mainly due to instabilities in the
St.~Ana ECR ion source. The spikes in current for the Faraday cup at the
target location following its insertion and retraction are thought
to be due to current induced by the motion of the cup apparatus. 

\begin{figure}[ht]
\begin{center}
\includegraphics[width=0.8\columnwidth,angle=0]{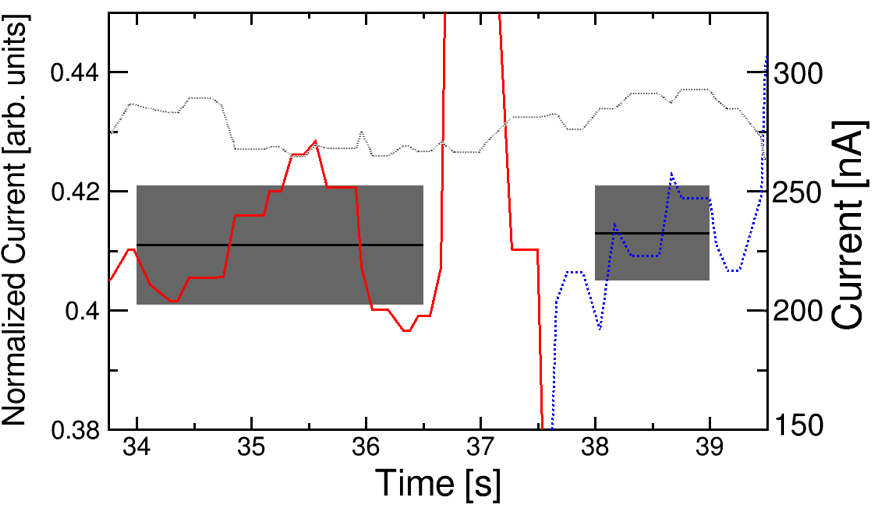}
\caption{
  (color online.) Demonstration of $>$99\% beam transmission through
  St.~George for 2.16~MeV $^{16}\rm{O}^{2+}$
  ions with the recoil separator tuned for $\rm{B}\rho=0.421$~Tm,
  $\rm{E}\rho=2.0$~MV (2~MeV ions, in this case.).
  The ratio (left axis) of the current on the target-cup (red line)
  and detector cup (blue-dashed line) to the upstream collimator
  current are shown for a time of near-constant upstream collimator
  current (gray-dotted line, right axis). Two gray boxes indicate
  (1) horizontally: the time-frame during which the target cup is
  not moving, in order to
  avoid unwanted current induced by the motion of the
  cup and (2) vertically: the average current ratio (black line) and
  standard deviation of the current ratio (gray band). 
  The mean current ratios differ by less than 1\%, indicating better
  than 99\% beam transmission.
\label{TransmissionUncertaintyDemo}}
\end{center}
\end{figure}

As seen in Figure~\ref{TransmissionUncertaintyDemo}, we observe full
transmission to a precision of $\sim$1~nA for a beam intensity of
$\sim$100~nA, i.e. $>99$\%.

\section{Results and Discussion}
\label{Results}
Energy acceptance measurements were performed as described in the
previous section for 8 ion beams, which were chosen to
span the phase space of $\rm{E}\rho$ and $\rm{B}\rho$ 
where recoil nuclei from $(\alpha,\gamma)$
reactions of astrophysical interest are anticipated. 
Table~\ref{BeamsMeasured} lists the ion beams
for which an energy acceptance measurement was performed.

\setlength{\tabcolsep}{2pt}
\begin{table}[ht]
\begin{center}
  \caption{\label{BeamsMeasured}
   Ion beams for which $\pm$8\% energy acceptance was verified for
	the St.~George recoil separator. Uncertainties for kinetic
	energy $T$, magnetic rigidity $B\rho$, and electric rigidity
	$E\rho$ are $\approx1$\%, which is dominated by the present uncertainty
	of the analyzing magnet energy calibration.}
	\def\arraystretch{1.25}
	\begin{tabular}{lccccc}
	\hline
	Species & Mass [u] & Charge [e] & T [MeV] & $\rm{B}\rho$ [Tm] & $\rm{E}\rho$ [MV] \\ \hline
	$^{16}\rm{O}^{4+}$ & 16 & 4 & 2.0 & 0.211 & 1.0 \\
	$^{1}\rm{H}^{1+}$ & 1 & 1 & 1.0 & 0.149 & 2.0 \\
	$^{4}\rm{He}^{1+}$ & 4 & 1 & 1.0 & 0.298 & 2.0 \\
	$^{20}\rm{Ne}^{4+}$ & 20 & 4 & 4.0 & 0.333 & 2.0 \\
	$^{16}\rm{O}^{2+}$ & 16 & 2 & 2.0 & 0.421 & 2.0 \\
	$^{4}\rm{He}^{2+}$ & 4 & 2 & 3.0 & 0.258 & 3.0 \\
	$^{16}\rm{O}^{4+}$ & 16 & 4 & 6.0 & 0.365 & 3.0 \\
	$^{16}\rm{O}^{4+}$ & 16 & 4 & 8.0 & 0.421 & 4.0 \\
	\hline
	\end{tabular}
\end{center}
\end{table}


\begin{figure}[ht]
\begin{center}
\includegraphics[width=1.0\columnwidth,angle=0]{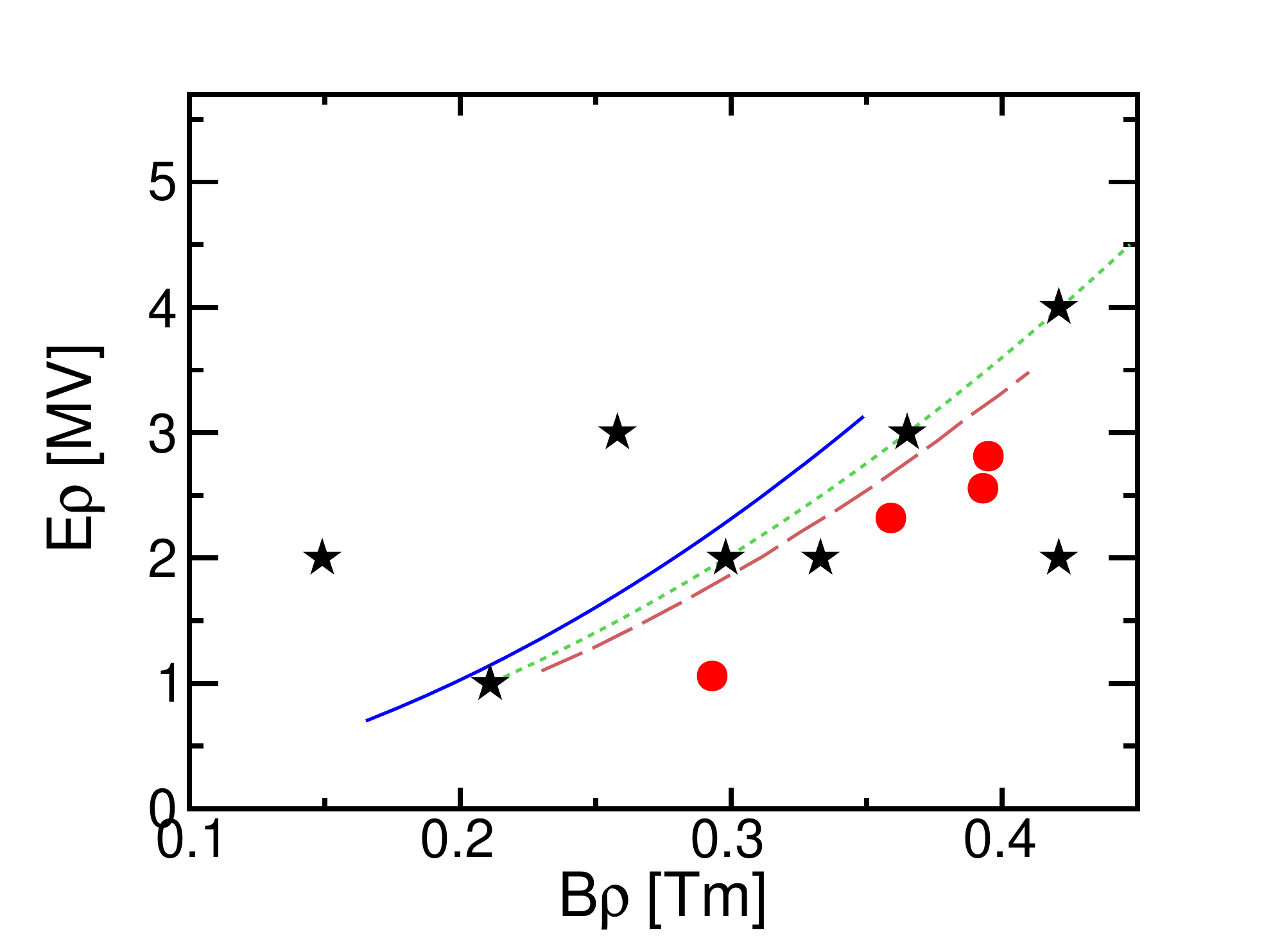}
\caption{
(color online.) Rigidity phase space designed to be accepted by the St.~George
recoil separator, where the $\rm{B}\rho$--$\rm{E}\rho$ 
of ions with a measured energy acceptance are indicated with black stars.
Examples of recoils of astrophysical interest, low recoil-energy
cases from Table~1 of
Reference~\cite{Coud08}, assuming the most populous equilibrium
charge-state from the dilute gas model of Reference~\cite{Saye77}
that is within the $\rm{B}\rho$--$\rm{E}\rho$ limits of St.~George,
are indicated by the red-filled circles. The solid blue line indicates the
$\rm{B}\rho$--$\rm{E}\rho$ space for $q=2^{+}$ recoils from the
$^{3}\rm{He}(\alpha,\gamma)^{7}\rm{Be}$ reaction at the energies measured
in the past with the recoil separator ERNA~\cite{Dile09}.
The long-dashed mauve line corresponds to $^{26}\rm{Mg}^{6+}$
recoils from the $^{22}\rm{Ne}(\alpha,\gamma)$ reaction for energies
covered by a previous measurement in forward
kinematics~\cite{Wolk89}.
The short-dashed green line corresponds to $^{16}\rm{O}^{4+}$ recoils from
$^{12}\rm{C}(\alpha,\gamma)$ for center of mass energies from
700--3000~keV.
\label{RigidityMap}}
\end{center}
\end{figure}

The field strengths for the ion-optical elements of St.~George which
were used for the energy acceptance measurements generally agreed
with the values provided by the calculations featured in
Reference~\cite{Coud08}. However, the absence of an angular opening
of the beam for the energy acceptance measurements performed here
allowed for considerable flexibility in the ion optical tune of
St.~George. We expect the full angular and energy acceptance
measurements performed in the future to provide much tighter
constraints on the settings of the ion-optical elements.


We compare the magnetic rigidity of measured ions to recoils of
astrophysical interest in Figure~\ref{RigidityMap}. Our measurements
verify the required energy acceptance for the majority of the range
of astrophysical interest, where example cases considered for the
design of St.~George (See 
Table~1 of Reference~\cite{Coud08}.) are chosen for comparison.
We show only the low $\rm{E}\rho$ cases highlighted in
Reference~\cite{Coud08}, since the
high $\rm{E}\rho$ cases are in general less astrophysically
interesting due to the corresponding high kinetic energies that are
absent in all but the highest temperature environments.
Our acceptance measurements also demonstrate the
capability to, for instance, repeat the
$^{3}\rm{He}(\alpha,\gamma)^{7}\rm{Be}$ measurement performed with
the recoil separator ERNA~\cite{Dile09}. Such a measurement may help to
improve constraints on Big Bang nucleosynthesis~\cite{Cybu16} and
resolve discrepancies in solar fusion calculations~\cite{Adel11},
particularly if performed in conjunction with angle-resolved $\gamma$-detection at
the target location. Furthermore, a direct measurement of
$^{22}\rm{Ne}(\alpha,\gamma)^{26}\rm{Mg}$ in the energy range
previously studied in regular kinematics~\cite{Wolk89} is well
within reach, which would help constrain the role of the
$^{22}\rm{Ne}(\alpha,n)$ $s$-process neutron source. Finally, we
demonstrate our ability to accept $^{16}$O recoils from
$^{12}\rm{C}(\alpha,\gamma)$, noting that the large angular
acceptance anticipated for St.~George will allow the lowest-energy
direct measurement of this reaction to date.

We used the measured width of the beam at the mass slits, as
determined in Step~1.7, to provide a first-order estimate of the
anticipated beam rejection capabilities of St.~George. 
The smallest
anticipated spatial separation between beam and recoil nuclei
provided by the Wien filter is $\approx$5~cm, which is for the case of
an $(\alpha,\gamma)$ reaction on an $A=40$ nucleus. In order to achieve the
desired mass rejection of 1 part in $10^{15}$, 
the magnitudes of the beam and recoil nuclei distributions must be
similar at the location of the mass slits when the amplitude of the
beam distribution is $10^{15}$ times larger than the amplitude of
the recoil distribution.
If we assume beam and recoil nuclei both have Gaussian horizontal distributions
with the same width, but the beam distribution is $10^{15}$ larger
in amplitude and has a centroid 5~cm horizontally displaced from the
recoil distribution centroid, a standard deviation of
$\sigma$$\approx$4~mm for both
distributions is sufficient to achieve a 1:1 ratio of beam-to-recoil
nuclei -3$\sigma$ left of the beam-line center at the mass slits
(3$\sigma$ is chosen as the point for comparison since $\pm3\sigma$
encompasses 99.8\% of the recoil distribution.).
In Step~1.7 we determined the width of the ion distribution at the mass
slits is $w\leq20$~mm, which corresponds to intercepting
$\lesssim$1~pA out of the full $\sim$100~nA ion intensity (i.e.
$\sim$0.001\%). For a
Gaussian distribution, this implies a mass slit opening of
$\pm$4.5$\sigma$. Therefore, from our measurement of $w$, we deduce
$\sigma\lesssim$2.2~mm, which is far less than the 4~mm required for
the desired mass rejection. We note that our estimate is a best-case
scenario since more pathological beam distributions
are often observed\footnote{This has been observed by
an in-beam fluorescing quartz viewer located after the mass slits,
which was added to the set-up following the measurements described
in this work.}. Furthermore, this estimate
neglects effects such as beam scattering within the separator which
adversely impact the mass rejection.

\section{Conclusions}
\label{Conclusions}
We present first results from commissioning of the
St.~George recoil separator. By performing measurements of stable
ion beams of varied elements, charge-states, and energies, we verify
that St.~George meets the design specifications of providing an
energy acceptance of $\pm8$\% within the separator's rigidity phase
space. Our ultimately employed magnetic field strengths generally
agree with predictions from ion-optical calculations performed with
{\tt COSY Infinity}; however, some deviations persist.
We attribute present disagreements to the
absence of realistic field maps in the calculations and an absence of
an angular opening of the beam at the target location, which will be
more restrictive when determining the optimum ion-optical tune. Our comparison
to the properties of recoils from anticipated future nuclear
reaction measurements demonstrates that St.~George will be a vital
tool used to obtain precise cross sections for $(\alpha,\gamma)$ 
reactions of astrophysical interest.

\section*{Acknowledgements}
\label{Acknowledgements}
We thank the staff of the Nuclear Science Laboratory at the University of Notre Dame for their outstanding support.
This material is based upon work supported by the National Science
Foundation under Grants No. 1062819, 1419765, and 1430152, and the
Nuclear Regulatory Commission under Grant No. NRC-HQ-12-G-38-0073.




\bibliographystyle{model1a-num-names}
\bibliography{StGeorgeReferences}





\end{document}